\documentclass[12pt,preprint]{aastex}

\begin{document}

\newcommand{\feh}{[{\rm Fe/H}]}
\newcommand{\znh}{[{\rm Zn/H}]}
\newcommand{\nstrs}{10~}
\newcommand{\msol}{M_\odot}
\newcommand{\pkts}{P_{KS}}
\newcommand{\etal}{et al.\ }
\newcommand{\delv}{\Delta v}
\newcommand{\Lya}{Ly$\alpha$ }
\newcommand{\lya}{Ly$\alpha$ }
\newcommand{\ndmp}{31 }
\newcommand{\nnew}{14 }
\newcommand{\nrun}{8,500}
\newcommand{\kms}{km~s$^{-1}$ }
\newcommand{\cm}[1]{\, {\rm cm^{#1}}}
\newcommand{\N}[1]{{N({\rm #1})}}
\newcommand{\e}[1]{{\epsilon({\rm #1})}}
\newcommand{\f}[1]{{f_{\rm #1}}}
\newcommand{\rAA}{{\AA \enskip}}
\newcommand{\sci}[1]{{\rm \; \times \; 10^{#1}}}
\newcommand{\ltk}{\left [ \,}
\newcommand{\ltp}{\left ( \,}
\newcommand{\ltb}{\left \{ \,}
\newcommand{\rtk}{\, \right  ] }
\newcommand{\rtp}{\, \right  ) }
\newcommand{\rtb}{\, \right \} }
\newcommand{\ohf}{{1 \over 2}}
\newcommand{\nohf}{{-1 \over 2}}
\newcommand{\rhf}{{3 \over 2}}
\newcommand{\smm}{\sum\limits}
\newcommand{\perd}{\;\;\; .}
\newcommand{\cmma}{\;\;\; ,}
\newcommand{\intl}{\int\limits}
\newcommand{\mkms}{{\rm \; km\;s^{-1}}}
\newcommand{\ew}{W_\lambda}
\def\ltsima{$\; \buildrel < \over \sim \;$}
\def\simlt{\lower.5ex\hbox{\ltsima}}
\def\gtsima{$\; \buildrel > \over \sim \;$}
\def\simgt{\lower.5ex\hbox{\gtsima}}

\title{Constraints on the origin of Manganese from the composition of the 
Sagittarius Dwarf Spheroidal Galaxy and the Galactic Bulge\footnote{This
paper was accepted for publication in the Astrophysical Journal Letters in 
June, 2003.}}

\author{ANDREW McWILLIAM}
\affil{The Observatories of the Carnegie Institute of Washington}
\affil{813 Santa Barbara St. \\
Pasadena, CA 91101.\\
email: andy@ociw.edu\\
\quad }

\author{R. MICHAEL RICH}
\affil{University of California, Los Angeles}
\affil{Department of Physics \& Astronomy\\
Math-Sciences 8979\\
Los Angeles, CA 90095.\\
email: rmr@astro.ucla.edu\\
\quad}

\and 

\author{TAMMY A. SMECKER-HANE}
\affil{Department of Physics \& Astronomy}
\affil{4129 Frederick Reines Hall\\
University of California \\
Irvine, CA 92697-4575.\\
email: tsmecker@uci.edu}

\begin{abstract}

The trend of [Mn/Fe] in the Galactic bulge follows the solar-neighborhood
relation, but most stars in the Sagittarius dwarf spheroidal galaxy 
(Sgr dSph) show [Mn/Fe] deficient by $\sim$0.2 dex.  This leads us to 
conclude that the Mn yields from both type~Ia and type~II SNe are 
metallicity-dependent.  Our observations militate against the idea, 
suggested by Gratton, that Mn is over-produced by type~Ia SNe, relative 
to type~II SNe.  We predict Mn/Fe ratios, lower than the solar 
neighborhood relation, for the younger populations of nearly all dwarf 
galaxies, and that Mn/Fe ratios may be useful for tracing the accretion 
of low-mass satellites into the Milky Way.

\end{abstract}



\section{Introduction}

In this Letter, we consider the abundance trend of Mn in three contrasting 
stellar populations: a sample of Galactic bulge K giants,
giants from the Sgr dwarf spheroidal galaxy, and stars from the
solar neighborhood.  

The bulge population formed rapidly (Ortolani et al. 1995), on a timescale 
less than a few Gyr, as evident in the trends of alpha-element
(e.g. O, Mg, Si, Ca, Ti) to iron ratios (McWilliam \& Rich 1994; 
Rich \& McWilliam 2000).  Chemical evolution models of these abundance trends 
support an even more rapid enrichment timescale for the bulge, $\leq$1~Gyr  
(e.g. Matteucci et al. 1999).  

Most dwarf spheroidal
galaxies have extended star formation histories, lasting many Gyr (e.g. Mateo 1998, 
Grebel 2000), as reflected by their sub-solar trends of $\alpha$/Fe
and metallicity spread (Shetrone et al. 2001, Bonifacio et al. 2000 and
Smecker-Hane \& McWilliam 2003, hereafter SM03).
The range of iron abundance for stars in the Sgr dSph exceeds a factor of ten,
$-1.5 \leq $ [Fe/H] $ \leq 0$ dex (e.g. SM03).  Thus, the Sgr dSph and Galactic bulge iron
abundance ranges are nearly the same, which makes these systems ideal to compare and contrast.


Arnett(1971) predicted that Mn yields from type~II SNe depend on the
neutron excess, and so should be metallicity dependent.
Woosley \& Weaver (1995) computed
yields for type~II SNe, covering a range of masses and metallicity; their Mn/Fe
yield ratios decreased by a factor of 3 from solar to one tenth solar metallicity,
but below Z=0.1Z$_{\rm \odot}$ the predicted yield ratio was constant.

Deficiencies of Mn, relative to Fe, in metal-poor stars were first noted
by Wallerstein (1962), consistent with the odd-even effect for
iron-peak elements suggested by Helfer et al (1959).
G89 measured Mn in a sample of disk and halo stars, and found a 
roughly constant [Mn/Fe]$\approx$$-$0.4 dex for [Fe/H]$<$$-$1, with the [Mn/Fe]
ratio increasing linearly for [Fe/H]$>$$-$1, up to the solar [Mn/Fe] value.

Gratton noted that this behavior is opposite to the trend of
   alpha elements with [Fe/H].  In that case, the
   [$\alpha$/Fe] ratio increases with decreasing [Fe/H], and reaches a
   constant value of [$\alpha$/Fe]$\sim$+0.4 dex below [Fe/H]=$-$1. 
   Tinsley (1979) explained this as due
   to the change in the ratio of type~II to type~Ia material, incorporated into stars
   at a given metallicity.
   Type~II SNe arise from massive stars of initial mass M$\simgt$10M$_{\odot}$, whose main-sequence
   lifetimes are a $\simlt$$10^{8}$ years.  A type~Ia SN arises in
   a binary system from mass transfer onto a white dwarf star (e.g. Iben \& Tutukov 1987, 
   Tornamb\`e \& Matteucci 1986),
   which explodes when
   the Chandrasekhar limit is exceeded; explosion times vary from $\sim$10$^8$ yr to $\geq$ 10~Gyr 
   after the stars were formed (e.g. Smecker-Hane \& Wyse 1992).

   In Tinsley's model, at low metallicity and early times stars were made from ejecta
   of type~II SNe, rich in oxygen and other alpha elements. 
   Type~Ia SNe did not become significant sources of nucleosynthesis products
   until after $\sim$1~Gyr when the metallicity had reached [Fe/H]$\sim$$-$1; at this
   point the low [$\alpha$/Fe] ratios from type~Ia SNe progressively reduced the
   ambient value until the solar composition was reached $\approx$ 4.5~Gyr ago.

G89 suggested that over-production of manganese in type~Ia SNe could account
for the observed [Mn/Fe] trend using Tinsley's paradigm;
indeed a bi-valued Mn/Fe yield ratio is perhaps the simplest assumption to make, based on
the early observations.  

Figure~1 summarizes the [Mn/Fe] values from recent studies of
Galactic stars for [Fe/H]$\simgt$$-$2.5; it shows a $\sim$0.5 dex increase
in [Mn/Fe] from [Fe/H]=$-$1.3 to $+$0.4, roughly linear with metallicity.
The Feltzing \& Gustafsson (1998, hereafter FG98),
Prochaska \& McWilliam (2000, hereafter PM00\footnote{The PM00 data are the Nissen et al. (2000)
results with improved $hfs$ corrections applied.}) and Reddy et al. (2003, hereafter R03)
data have been put onto the same scale by applying corrections to ensure
solar [Mn/Fe] in the interval $-$0.10$\leq$[Fe/H]$\leq$$+$0.10.  This normalization
is not possible for the remaining studies shown in Figure~1; for those data we expect
zero-point uncertainties near $\sim$0.10 dex.

The FG98 results for metal-rich stars show [Mn/Fe]
increasing at higher [Fe/H], whereas the [$\alpha$/Fe] ratios remain constant;
this suggests a metallicity-dependent yield for Mn produced by type~Ia SNe,
not Mn over-production by type~Ia SNe.

Nissen \& Schuster (1997) showed that thick
disk stars possess $\alpha$-element enhancements independent of [Fe/H], yet
the thick-disk stars in Figure~1 show a clear trend of increasing [Mn/Fe] with 
[Fe/H]; this can be understood with a metallicity-dependent Mn yield from type~II SNe.

The PM00/N00 data display a step in [Mn/Fe],
near [Fe/H]=$-$0.7, which may be associated with the transition from thin to thick disk
populations, apparently confirmed by the slope of the R03 data.
If real, the putative plateau might reasonably be explained by Mn over-production in
type~Ia SNe, or by a metallicity-dependent yield combined with the large metallicity
dispersion in the Galactic disk age-metallicity relation. 
However, possible systematic errors that we suspected in PM00, and the lack of evidence for the 
step in the data of P00 and G89, cast doubt on the reality of this feature.

\section{Observations, Reductions and Analysis}

High resolution (R=34,000--50,000) spectra of individual stars in both 
the Galactic bulge, through Baade's Window, and the Sgr dSph were obtained using the 
Keck~I\footnote{The W.M. Keck Observatory, is operated as a scientific partnership among
the California Institute of Technology, the University of California, and the
National Aeronautics and Space Administration.  The Observatory was made possible by
the generous financial support of the W.M. Keck Foundation.  We extend special thanks
to the people of Hawaiian ancestry on whose sacred mountain we were privileged to be guests.}
telescope and echelle spectrograph (HIRES, Vogt et al. 1994); the typical S/N ratios were 50
per extracted pixel.  Complete details of these two investigations can be found in
SM03 and McWilliam \& Rich (2003, in preparation).  The Sgr dSph
spectra were extracted using the IRAF suite of routines, whereas
the bulge star spectra were extracted with MAKEE, written by T. Barlow.
Line equivalent widths were measured using a semi-automated routine GETJOB
(McWilliam et al 1995).  From 1 to 4 Mn~I lines were measured for each of the
Galactic bulge giants, typically 3 lines.  For the Sgr dSph from 4 to 10 Mn~I 
lines were measured per star, most often 9 lines.
Abundances were computed from the equivalent widths
using the spectrum synthesis program MOOG (Sneden 1973) and the 64-layer Kurucz
(1993) model atmospheres.  The model atmosphere parameters were chosen, using
photometric and spectroscopic methods, detailed in 
McWilliam \& Rich (2003) and SM03.  Since Mn~I
lines are strongly affected by hyperfine splitting, we employed
line lists generated from published $hfs$ constants or taken from Kurucz (1997) 
for the hyperfine components of each line, and then performed the appropriate 
synthesis to derive Mn abundances.

\section{Results and Discussion}

\subsection{The Galactic Bulge}

A plot of [Mn/Fe] versus [Fe/H] for the Galactic bulge stars is presented in Figure~2.
The present data indicate that the bulge [Mn/Fe] trend is approximately the same as
found fo solar neighborhood disk and halo stars, and even follows the local [Mn/Fe]
trend for metal-rich stars.  

We note that the bulge results are based on analysis of red giant stars, whereas
the solar neighborhood points are almost exclusively from dwarf or turn-off stars;
this may introduce zero-point abundance differences, expected to be less than 0.10 dex
in [Mn/Fe].  To investigate zero-point abundance uncertainties for red giants
we measured the [Mn/Fe] and [Fe/H] ratios for the thick-disk red giant Arcturus, using
the Hinkle et al. (2000) spectrum.  We found [Mn/Fe]=$-$0.23 and [Fe/H]=$-$0.56 for Arcturus, 
0.075 dex below the mean [Mn/Fe] for the thick-disk dwarf stars of P00,
in the range $-$0.63$\leq$[Fe/H]$\leq$$-$0.50.  If the zero-point of the bulge [Mn/Fe] 
is increased by the 0.075 dex shift for Arcturus, agreement between bulge and solar neighborhood 
[Mn/Fe] is improved.  We adopt the 0.075 dex zero-point shift as a measure of the systematic 
uncertainty of the red giant [Mn/Fe] abundance ratios in this letter; [Fe/H] for the red giants 
are accurate at the 0.10 dex level.

%
%

Figure~2 is inconsistent with the G89 scenario of Mn over-production by type~Ia SNe.
The high [$\alpha$/Fe] and [Eu/Fe] ratios observed in bulge stars (McWilliam \& Rich 1994, 
McWilliam \& Rich 1999, Rich \& McWilliam 2000) suggest nucleosynthesis in the bulge was 
dominated by type~II SNe with a rapid formation time scale (e.g. Matteucci et al. 1999), with 
element ratios similar to the halo but at higher overall metallicity.  Thus, if type~Ia SNe 
over-produced Mn then bulge stars would show a deficiency in [Mn/Fe] ratios; but such a
deficiency is not observed.  The [Mn/Fe] trend seen in Figure~2 is entirely
consistent with a metallicity-dependent yield for Mn in type~II SNe.  

If type~II SNe produce a metallicity-dependent Mn yield above [Fe/H]=$-$1, then
the linear trend of [Mn/Fe] with [Fe/H] for the Galactic bulge stars in Figure~2
is consistent with evolution characterized by rapid recycling of the gas, as
in the Simple Model (Searle \& Sargent 1972), which assumed a closed box 
and instantaneous chemical recycling.  Rich (1990) first showed that
the bulge iron abundance distribution can be fit by the Simple Model, and this is
confirmed by Zoccali et al. (2003) using iron abundances inferred from $V-K$ photometry.
Thus, the bulge [Mn/Fe] trend with [Fe/H] supports the metallicity-dependent Mn yield for
type~II SNe (above [Fe/H]=$-$1), suggested by Arnett (1971) and Woosley \& Weaver (1995).
More data would be helpful for confirming these conclusions; in this regard it
would be useful to measure the [Mn/Fe] trend in bulge stars well below [Fe/H]$\sim$$-$1.

\subsection{The Sagittarius Dwarf Spheroidal Galaxy}

In Figure~3, we present the results for [Mn/Fe] in the Sgr dSph.
The most obvious feature is that the metal-rich Sgr dSphs stars
with [Fe/H] $\simgt -0.6$ have [Mn/Fe] values that lie well below,
by 0.2 dex, those of solar neighborhood stars.


The Sgr dSph and Galactic bulge abundances were computed with spectra of similar quality,
with the same atomic parameters, model atmosphere grid, and temperature scale. 
Systematic zero-point differences and random errors are unable to explain the difference 
between our Sgr and bulge [Mn/Fe] ratios.


The unusually low [Mn/Fe] ratios in Sgr dSph are inconsistent with Mn over-production
in type~Ia SNe, suggested by G89.  The low [$\alpha$/Fe] ratios 
(see SM03) strongly suggest that type~Ia SNe dominated
the nucleosynthesis of iron-peak elements in this galaxy for 
[Fe/H]$\simgt$$-$0.6 dex.  Given an increased nucleosynthetic contribution from type~Ia
SNe in the Sgr dSph, [Mn/Fe] ratios enhanced over the solar neighborhood ratios are expected
if Gratton's hypothesis
is correct.  Our observations of {\it deficient} [Mn/Fe] ratios in Sgr dSph stars are
opposite of this expectation.

Observations of s-process neutron capture elements in the Sgr dSph show that ejecta from
the old, metal-poor, AGB population dominated the chemical enrichment of the 
neutron-capture elements, in the epoch probed by stars with [Fe/H]$\simgt$$-$0.6 (SM03).  
%
%
In this paradigm for the Sgr dSph the old, metal-poor, AGB stars must have been
accompanied by low-metallicity type~Ia SNe, which we assume 
dominated iron production prior to the formation of the [Fe/H]$\ge$$-$0.6 population.
This picture of the chemical evolution of the Sgr dSph provides a natural explanation
for the observed low [Mn/Fe] values, if the Mn/Fe yield ratio increases with metallicity
in type~Ia SNe: the low [Mn/Fe] ratios reflect the fact that iron-peak nucleosynthesis
was dominated by the metal-poor type~Ia SNe population.  

If all Sgr dSph stars with [Fe/H]$\ge$$-$0.6 possess the same [Mn/Fe] ratio, then
it must be that element yields from low-metallicity SNe ([Fe/H]$\le$$-$1) completely
overwhelmed products from more metal-rich stars.  On the other hand,
an upward slope of [Mn/Fe] with [Fe/H] would indicate a detectable contribution from 
metal-rich type~Ia and type~II SNe, albeit  dwarfed by the metal-poor type~Ia component.
More data is required to distinguish between these two possibilities.
In either situation the current data argue that the old, metal-poor, population produced much 
more iron-peak enrichment through type~Ia SNe than the younger, metal-rich,
populations produced via type~Ia and type~II SNe.

Since the Mn/Fe ratios in Sgr are the consequence of slow star formation and significant mass loss,
other dwarf spheroidal systems with extended star formation histories should exhibit
similar abundance trends.  In this regard it would be very useful to study the trend of [Mn/Fe] 
in the Magellanic Clouds and Local group dSph systems, such as the Fornax dSph.  We note that
sub-populations in the
bright Galactic globular cluster $\omega$~Cen possess some chemical similarities to the
Sgr dSph.  Deficient [Cu/Fe] ratios in $\omega$~Cen (Cunha et al. 2002) may indicate
a chemical enrichment history similar to Sgr dSph, but at lower overall [Fe/H].  We predict
that the most metal-rich stars in $\omega$~Cen will show deficient [Mn/Fe] ratios, and stars in
Sgr~dSph possess low [Cu/Fe] ratios.

The chemical enrichment model proposed by SM03 for Sgr dSph involves delayed injection of
nucleosynthesis products from an old, metal-poor, population and suggests a relatively
high ratio of metal-poor to metal-rich stars.   This condition is expected
of galaxies which lose a significant fraction of gas over long timescales
and galaxies which accrete pristine material sufficient to maintain a large metal-poor 
fraction.  The model
qualitatively explains the unusual chemical composition of the Sgr dSph stars, but no 
quantitative modeling has been undertaken; thus we cannot yet say what fraction of old,
metal-poor, stars is required to produce the observed abundance ratios.  
To quantitatively constrain the models we must first obtain unbiased age and [Fe/H] 
distributions for Sgr dSph stars.
%

Even near solar metallicity stars in the Sgr dwarf have [Mn/Fe] significantly lower
(by $\sim$0.2 dex) than stars in the disk/bulge.  Thus, if Sgr-like systems
fell into the bulge and were so thoroughly mixed that no trace of their distinct
kinematics remained, it would still be possible to infer their presence through the
depressed [Mn/Fe] and [$\alpha$/Fe] ratios, and enhanced s-process abundances.


\section{Summary}

We report on the abundance trends of manganese in the Galactic bulge and
the Sagittarius dwarf spheroidal galaxy.  Both stellar populations show
a general trend of [Mn/Fe] increasing with higher [Fe/H];
but while the bulge follows roughly the solar neighborhood [Mn/Fe] trend, the 
trend in Sgr is lower by $\sim$0.2 dex.
We believe that this offset reflects the less rapid chemical evolution of the
Sgr dwarf compared to the Galactic bulge and solar neighborhood.  Similar trends should be
present in all dwarf galaxies, where chemical enrichment over long time scales reached 
metallicities greater than [Fe/H]=$-$1.  The low Mn abundance at a given [Fe/H]
could be used to identify stars from accreted dwarf galaxies 
in large-scale surveys of the disk and bulge.

The [Mn/Fe] trend in the Galactic bulge suggests a metallicity-dependent yield of Mn in
type~II SNe, qualitatively consistent with the predictions of Arnett (1971) and
Woosley \& Weaver (1995).  A metallicity-dependent Mn yield from type~II SN is supported by
the trend of [Mn/Fe] in local thick disk stars which are known to have constant,
enhanced [$\alpha$/Fe].  Given the rapid bulge formation timescale (Matteucci et al. 1999), the 
metallicity dependence of the Mn/Fe yield ratio is likely very similar to the observed trend;  
in this regard it will be interesting to measure [Mn/Fe] in metal-poor Galactic bulge stars.

The deficient [Mn/Fe] ratios in the Sagittarius dwarf spheroidal galaxy
implies a metallicity-dependent Mn yield for type~Ia SNe; 
this is supported by the observations of FG98, which show increasing [Mn/Fe] in metal-rich disk
stars, at constant [$\alpha$/Fe].  The idea that Mn is over-produced by type~Ia SNe, as
suggested by G89, is ruled-out by these observations.

One complication in our interpretation is the apparent offset in [Mn/Fe] between thin and
thick disk stars, which appears consistent with Mn overproduction in type~Ia SNe, but contrary
to the [Mn/Fe] trend in the metal-rich disk stars of FG98 and the Sgr dSph.  If this offset
is real it might be understood as the consequence of a metallicity-dependent yield combined
with a large metallicity-dispersion in the thin-disk age-metallicity relation.

We have insufficient information to state the form of the metallicity-dependent yield for
type~Ia SNe, other than it is lower at low [Fe/H].  Further understanding of Mn/Fe yields will
be made with detailed chemical evolution models for Sgr dSph, but this will require measurement
of the detailed age and [Fe/H] distribution functions for this galaxy.

\vskip0.2cm

\acknowledgments

{AM gratefully acknowledges support from NSF grants AST-96-18623 and
AST-00-98612, and would  like to thank the referee, B.~Carney and G.~W.~Preston.
RMR acknowledges support from AST-00-98612.
TSH acknowledges support from NSF grants AST-96-19460 and AST-00-70895,
and a AAS Small Research Grant.
}


\begin{figure}
\plotone{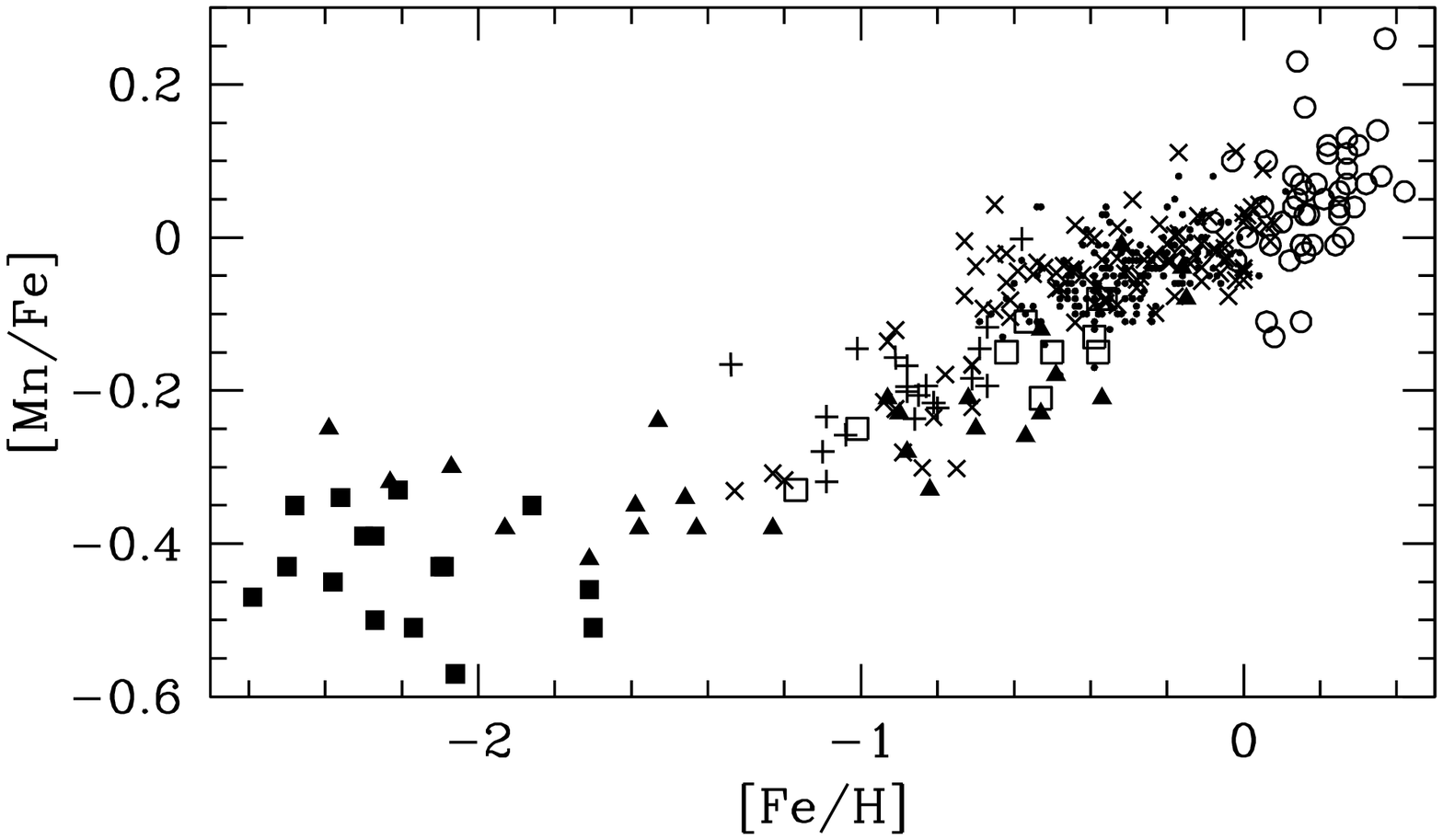}
\caption {\footnotesize [Mn/Fe] from numerous studies. Dots from R03,
crosses: PM00/N00, pluses: thick disk stars from PM00/N00,
open circles: FG98, triangles: G89, open squares: P00,
and filled squares: Johnson (2002).  Johnson (2002) analyzed red giants,
G89 studied a mix of dwarfs and giants;
all others employed dwarf or turnoff stars.   Zero-point 
corrections have been applied to the PM00/N00, R03 and FG98 data:
$+$0.02, $+$0.10, and $-$0.06 dex respectively.  General systematic errors 
in [Mn/Fe] are likely less than 0.10 dex, and random uncertainty of individual
points typically 0.06 dex; the [Fe/H] uncertainty is $\simlt$0.10 dex.
Johnson (2002) stars and G89 having stars [Fe/H]$\simlt$$-$1.2 are likely halo
objects.  G89 stars having $-$1$\leq$[Fe/H]$\simlt$$-$0.5 are likely thick disk members. }
\label{fig-mnfeall}
\end{figure}

\begin{figure}
\plotone{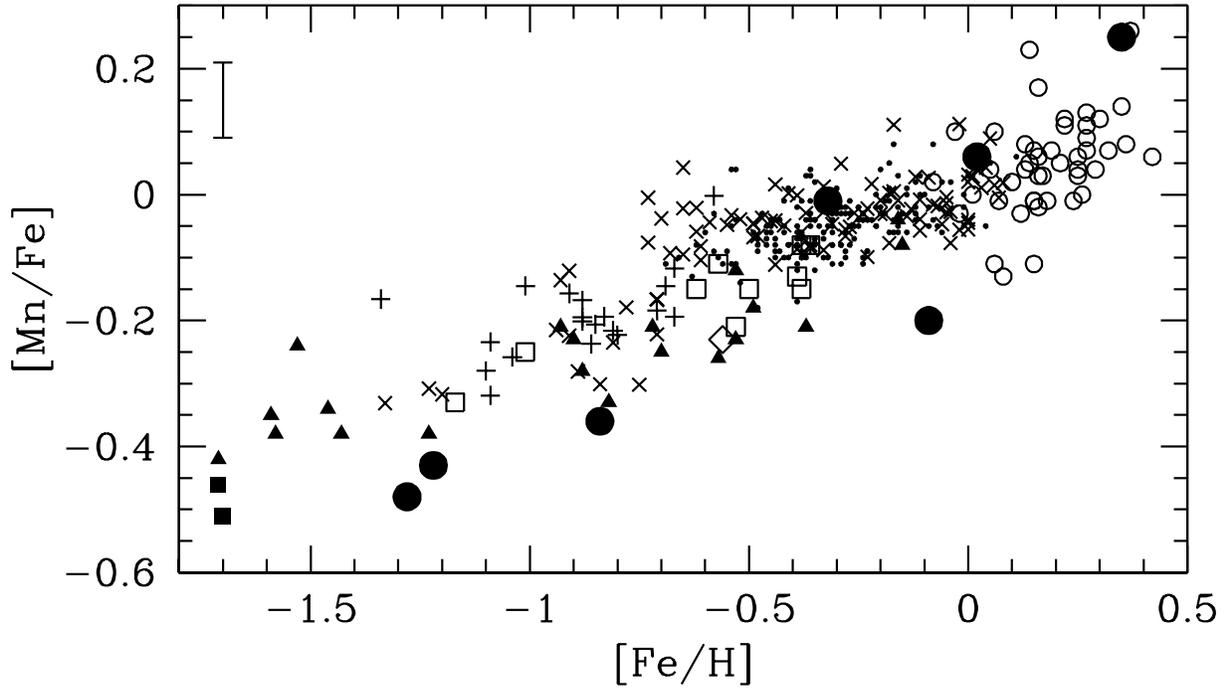}
\caption {\footnotesize Same as Fig.~1, but with bulge stars (filled circles) and 
Arcturus (open diamond) added.  The error bar indicates 1$\sigma$ scatter of [Mn/Fe] 
from Mn lines, not the error on the mean [Mn/Fe] value.  Although $\alpha$-elements 
and Eu are enhanced in bulge giants, Mn follows the Galactic disk relation.  This would 
not be the case if Mn were over-produced in type~Ia SNe.  }
\label{figure-bulge}
\end{figure}

\begin{figure}
\plotone{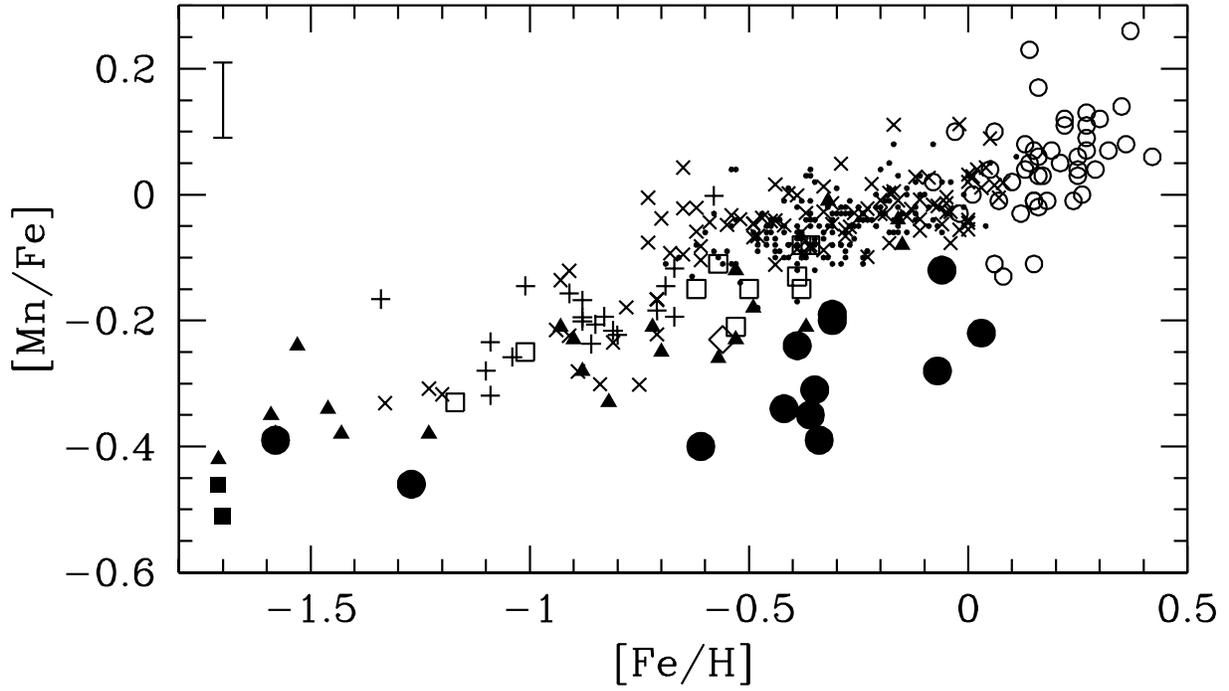}
\caption{\footnotesize Same as Fig.~1, but with Sgr dSph stars (filled circles) and 
Arcturus (open diamond) added.  The error bar indicates 1$\sigma$ scatter of [Mn/Fe] 
from Mn lines, not the error on the mean [Mn/Fe] value.  We propose that the low [Mn/Fe] 
ratios in Sgr dSph is a consequence of enrichment of the interstellar medium by metal-poor
type~Ia SNe.}
\label{figure-sgr}
\end{figure}


%
%
%
%
%
\end{document}